\documentclass[preprint,12pt]{elsarticle}

\usepackage{lineno,hyperref}
\usepackage{amssymb}
\usepackage[cmex10]{amsmath}
\usepackage{epstopdf}
\usepackage{algorithm}
\usepackage{textcomp}
\usepackage{blkarray, bigstrut} %
\usepackage{blindtext}
\usepackage[center]{caption}
\usepackage[font=bf]{caption} 
\usepackage{subcaption}
\usepackage{multirow}
\usepackage{mathtools}

\usepackage[a4paper,bindingoffset=0.2in,%
            left=0.9in,right=0.9in,top=0.9in,bottom=0.9in,%
            footskip=.25in]{geometry}

\usepackage[T1]{fontenc}
\usepackage[utf8]{inputenc}
\usepackage{tabularx,ragged2e,booktabs,caption}
\newcolumntype{C}[1]{>{\Centering}m{#1}}

\usepackage{algpseudocode}
\usepackage{array}

\journal{Optics Communications}

\begin{document}

\begin{frontmatter}

\title{Low-Complexity Detection of Multiweight Permutation Modulation Space-Time Block Codes for Indoor Visible Light Communication}
	
\author[1,2]{Oluwafemi Kolade}
\author[1]{Ling Cheng \corref{cor1}}
\ead{ling.cheng@wits.ac.za}

\cortext[cor1]{Corresponding author}
\address[1]{Optical Communication Laboratory, School of Electrical and Information Engineering, University of the Witwatersrand, Johannesburg, South Africa.}
\address[2]{Sibanye-Stillwater Digital Mining Laboratory (DigiMine), Wits Mining Institute (WMI), University of the Witwatersrand, Johannesburg, South Africa.}

\begin{abstract}
In this paper, the spectral efficiency of permutation modulation-based multiple input multiple output (MIMO) visible light communication is improved using systematically designed, multiweight codeword matrices. Soft-decision, low-complexity detection schemes are then designed for the receiver and compared with the maximum likelihood (ML) detection method. Bit error rate (BER) results show the soft-decision detection algorithm is able to decode the transmitted information without knowledge of the channel state information. This enables the mobile receiver decode information while within the field of view of the transmitter unit. The BER results also show a close match with the ML detection in some codebooks and the performance of the soft-decision decoder is evaluated for different positions of the receiver in an indoor environment.
\end{abstract}
\end{frontmatter}

\section{Introduction}
The concept of using existing illumination for Visible Light Communication (VLC) is becoming attractive for low-cost communication with potential to deliver information to the last mile. Since lighting infrastructures usually consist of multiple Light Emitting Diodes (LEDs), a natural Multiple Input Multiple Output (MIMO) setup is established. MIMO schemes provide several advantages such as higher data rates, spatial diversity and improvement of the error rate performance using the receiver diversity. For example, Space-Time Block Codes (STBC) \cite{tarokh-jafarkhani-calderbank} provide transmit diversity such as the Alamouti STBC \cite{alamouti}. The single receiver's Maximum Likelihood (ML) detection method requires the Channel State Information (CSI) but with reduced detection complexity. 

In MIMO VLC \cite{komine-nakagawa, mesleh-haas-ahn-yun, fath-haas}, activating a unique combination of LEDs from the total available LEDs can be used for information signaling. The index of the activated LED(s) can also convey additional information in order to increase the data rate. Repetition Coding (RC) is an example whose receiver adds up light intensities from LEDs transmitting the same information over the multiple transmitters at each transmit time. Index Modulation (IM) assists with improving the data rate of MIMO-VLC systems by mapping information to activated LEDs. 
In scenarios where Inter-Channel Interference (ICI) is required to be eliminated, Spatial Modulation (SM) \cite{mesleh-haas-ahn-yun}, a type of Index Modulation (IM) maps $\text{log}_2 M$ bits to an activated index. The activated index can also convey additional information selected from a conventional $M$-ary modulation scheme such as pulse amplitude modulation (PAM) or On-Off Keying (OOK) \cite{kahn-barry}. Another IM scheme is Space Shift Keying (SSK) \cite{jeganathan-ghrayeb-szczecinski-ceron} which activates one LED index at each transmit time in order to eliminate ICI with a receiver complexity similar to a Single Input Single Output (SISO) receiver complexity. Space-Time Shift Keying (STSK) proposed in \cite{suguira-chen-hanzo} extends the concept of Space Shift Keying (SSK) to take advantage of time diversity by activating one of the $Q$ dispersion matrices which have been prior mapped to the information bits. However, when MIMO detection schemes require the CSI, they become prone to channel estimation errors and impractical when the CSI is unknown. Differential Spatial Modulation (DSM) \cite{ishikawa-sugiura, bian-miaowen-xiang-poor-vincent-jiao-bingli, xiao-xiao-yang-liu-li-xiang, bayaki-schober} uses a permutation matrix to determine the antenna index to be activated and can also mitigate ICI between the transmitters since one transmitter is active within a transmit period. Also, the incoming information bits are differentially modulated, as a result the receiver is able to detect the transmitted signals without knowledge of the CSI.

Permutation Modulation (PM) \cite{slepian} is one of several methods of generating space-time codes used for antenna transmit matrices \cite{bian-miaowen-xiang-poor-vincent-jiao-bingli, suguira-chen-hanzo, kolade-cheng, lai-shih-lee-tu-chi-wu-huang, chi-yeh-lai-huang} and is preferred mainly because of its ability to mitigate the effect of ICI. In \cite{bian-miaowen-xiang-poor-vincent-jiao-bingli}, information bits are mapped on to PM-aided matrices which are then differentially modulated. In \cite{suguira-chen-hanzo}, the PM-aided matrices are used to design antenna dispersion matrices. The main drawback of PM-aided MIMO schemes is the reduced spectral efficiency when compared with conventional MIMO schemes such as RC, SM and generalized spatial modulation techniques \cite{jeganathan-ghrayeb-szczecinski, fath-haas}. Moreover, decoding PM-aided matrices becomes too complex as the number of antennas increase. A low-complexity decoder in \cite{kolade-versfeld-van-wyk} combines optimization algorithms in \cite{kuhn} and \cite{murty} to decode permutation block codes while a similar concept in \cite{kolade-cheng} decodes PM-aided transmit matrices at the receiver with and without the knowledge of the CSI at the receiver.
While the dispersion matrices are designed for Radio Frequency (RF) systems which can transmit either real or complex-valued signals, a PM STSK scheme suitable for VLC adopts the STSK scheme in \cite{suguira-chen-hanzo-stsk}. Similar to DSM, this scheme also activates a permutation matrix in which the non-zero elements in the matrix determine the LED index to be activated at each symbol period. %*** and a soft-decision decoder can reduce the decoder complexity and without knowledge of the CSI. ***

Optimization algorithms in \cite{kuhn} and \cite{murty} are then combined in \cite{kolade-cheng} in order to decode the received matrix with and without the knowledge of the CSI at the receiver. In an indoor environment where the receiver is mobile, it is of interest to have low-complexity detection algorithms that can operate without knowledge of the CSI and within practical complexity limits. On the other hand, permutation coded schemes are generally low-rate codes with limited low-complexity, soft-decision receivers. 
In order to increase the information rate of permutation-coded systems, concatenated permutation block codes \cite{heymann-weber-swart-ferreira} have been used while simultaneously increasing the code's minimum distance. However, the concatenated permutations presented in \cite{heymann-weber-swart-ferreira} do not construct equal weight matrices which is required for the MIMO scheme presented in this paper. On the other hand, soft-decision detection of permutation matrices with weight greater than 1 do not exist to the best of our knowledge.

In this paper, a novel higher rate PM-aided MIMO VLC scheme is presented at the transmitter and low-complexity detection schemes at the receiver. The first contribution of this paper introduces a class of concatenated permutation codes which generate equal weight space-time codes in order to increase the number of bits conveyed by the transmit matrix of the PM-aided MIMO VLC scheme. The second contribution designs low-complexity, iterative soft-decision (SD) detection methods using optimization algorithms. The SD methods are capable of decoding the likely transmitted bits without knowledge of the channel matrix. The BER expression for the ML decoder is also derived with the SD decoders matching the ML performance for some setups.

%In this paper, a soft-decision receiver of the permutation-aided MIMO VLC scheme in \cite{kolade-cheng} is designed using the Branch and Bound (B\&B) method with the assumption that the CSI is unknown at the receiver. This is informed by interpreting the output of the channel as an assignment problem \cite{kuhn, murty, kolade-versfeld-van-wyk, kolade-shimaponda-nawa-versfeld-cheng}. Then, we improve the data rate of the permutation-aided MIMO VLC scheme using the concatenated permutation sequences to create higher rate antenna transmit matrices. The soft-decision decoders are also designed to decode the received matrix assuming no knowledge of the CSI. Results show that while the data rate can be improved by generating more transmit matrices from concatenated codewords, the Bit Error Rate (BER) performance of the soft-decision decoder can match the ML performance for some setups. The computational complexities of the decoders are also analyzed. 

The following notations are used in this paper. tr$[\cdot]$ is the trace operation, $||\cdot||_F$ is the Frobenius norm, $(\cdot)^H$ is the conjugate transpose operation and the Q-function is defined as $\text{Q}(r) = \frac{1}{2\pi}\int_{r}^{+\infty} \text{exp}^{\frac{-x^2}{2}} \text{dx}$.

\section{System Model for Permutation-Aided MIMO For VLC}

\subsection{Permutation-Aided STSK Scheme}
Consider a permutation codebook $\boldsymbol{C}$ with $Q$ codewords. Each codeword $\boldsymbol{c}_q$ at row $q$ ($q = 1, 2, \dots, Q$) is a row vector consisting of a unique permutation of integers $(c_1,c_2, \dots, c_L)$. Hence, each symbol in a codeword creates a corresponding column vector
\begin{align*}
    (\underbrace{0,\dots,0}_{l-1},1, \underbrace{0, \dots, 0}_{L-l})^T,
\end{align*}
and a permutation of $L$ symbols in a codeword produces a codeword matrix $\boldsymbol{P}_q$. Each matrix $\boldsymbol{P}_q$ is equivalent to the codeword at row $q$ and the index of the non-zero element in each row of $\boldsymbol{P}_q$ corresponds to each integer in $\boldsymbol{c}_q$. %Each integer indicates the column with a 1 while other columns are 0. 
%in the permutation matrix. 

\begin{figure*}
  \centering\includegraphics[width=35pc]{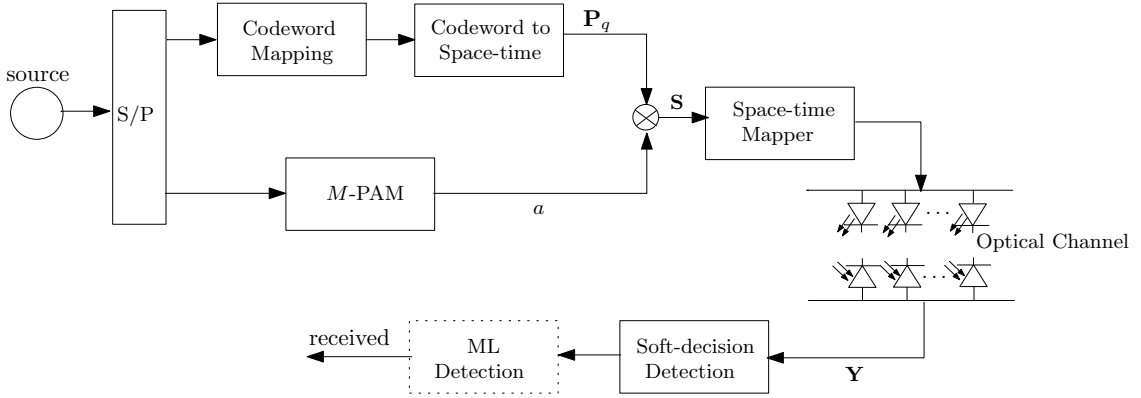}
    \caption{System model for permutation modulation-aided STSK over a VLC channel with soft-decision detection.}
    \label{fig:stsk-system-model}
\end{figure*}

We define the weight $w$ of the matrix as the number of non-zero elements in its rows and columns. Hence, a $\{0, 1\}^{L \times L}$ permutation matrix $\boldsymbol{P}_q$, having columns and rows with weight $w = 1$ is transmitted within a block of multiple time slots. 
%The non-zero element in each row in $\boldsymbol{P}_q$ corresponds to each integer in $\boldsymbol{c}_q$. Hence, each integer indicates the column with a 1 while other columns are 0. 
Considering the model in Fig. \ref{fig:stsk-system-model}, the incoming bits are mapped on to $\boldsymbol{P}_q$ and an $M$-PAM intensity level to create a transmit matrix. The permutation matrix $\boldsymbol{P}_q$ enables an activation pattern per transmit block, with each non-zero element representing the activated LED index. As a result, each transmit block conveys $\text{log}_2 Q$ bits from the incoming message bits. The power constraint of the transmitted block is given as $\text{tr}[\boldsymbol{P}^H_q \boldsymbol{P}_q]$.
Additional $\text{log}_2 M$ information bits, mapped onto a unipolar $M$-ary modulation scheme such as Pulse Amplitude Modulation (PAM) can be transmitted over the active LED in order to increase the information rate. This also ensures the transmitted signal is positive and real and a total of $\text{log}_2 (Q \cdot M)$ bits can be transmitted at each transmit block.
The resulting transmitted matrix $\boldsymbol{S} = [s_{ij}]$
\begin{equation} 
    \label{eq:modulation-times.permutation-matrix}
    \begin{aligned} 
        \boldsymbol{S} = a \cdot \boldsymbol{P}_q,
    \end{aligned}
\end{equation}
represents a space-time code where $a$ takes any of the $M$ intensities
\begin{equation} 
    \label{eq:pam-intensity}
    \begin{aligned} 
      I_m = \frac{2I}{w(M+1)}m \quad \text{for} \quad m = 1,2,\dots,M, %(n = 1, \dots, N),
    \end{aligned}
\end{equation}
where $I$ is the mean optical transmit power.
The $M$-PAM symbol is transmitted over each activated LED index such that
\begin{equation} 
    \label{eq:assignment-problem-explained}
    \begin{aligned} 
      \sum_{i=1}^{L} s_{ij} =  %(t = 1, \dots, T), \\
      \sum_{j=1}^{L} s_{ij} = a. %(n = 1, \dots, N),
    \end{aligned}
\end{equation}
As a result, the number of bits per symbol transmitted at each sampling period is given as $\frac{\text{log}_2(Q \cdot M)}{L}$.

\subsection{MIMO VLC Channel Model}
A line-of-sight (LOS) channel model between $L$ LEDs as transmitters and $L$ photodiodes (PD) as receivers are assumed. The transmitters employ intensity modulation while direct detection is employed at each receiver. Assuming Lambertian LEDs and PDs, the channel gain between an LED and a PD is given as \cite{gfeller-bpast}

\begin{equation}
h =
\begin{cases}
    \frac{(e+1)A_{\text{pd}}}{2\pi f^{2}} cos^e(\phi)cos(\psi), & \text{for } 0 \leq \psi \leq \Psi\\
    0,              & \psi > \Psi,
    \label{eq:vlc-tx-h}
\end{cases}
%h_{mn} =
%\begin{cases}
%    \sum_{k=1}^{K} \frac{A_{pd}}{f_{mn}^{2}} R_0(\psi)cos(\alpha_{mnk}), & \text{for } 0 \leq \alpha_{mnk} \leq \alpha_{c}\\
%    0,              & \alpha_{c} > \alpha_mnk},
%    \label{eq:vlc-tx-h}
%\end{cases}
\end{equation}
where $A_{\text{pd}}$ is the PD's surface area, $f$ is the distance between the LED and the PD, $\phi$ and $\psi$ describe the transmitter's and receiver's angle of incidence respectively. For a semi-angle $\Phi_{\frac{1}{2}}$, the Lambertian emission order $e = \frac{-\text{ln} 2}{\text{ln} (cos \Phi_{\frac{1}{2}} )}$, while $\Psi$ is the field of view of the receiver.
The received matrix at the photodiodes is modelled as
\begin{equation}
  \begin{aligned}
  \label{eq:received-stsk-rf-signal}
    \boldsymbol{Y} = \boldsymbol{H}\boldsymbol{S} + \boldsymbol{N},
  \end{aligned}
\end{equation}
where $\boldsymbol{Y} \in \mathbb{R} ^{L \times L}$ is the matrix received at each block, each element in $\boldsymbol{H} \in \mathbb{R} ^{L \times L}$ is given by the expression in ($\ref{eq:vlc-tx-h}$), $\boldsymbol{S} \in \mathbb{R} ^{L \times L}$ is the transmitted PM signal matrix and $\boldsymbol{N} \in \mathbb{R} ^{L \times L}$ consists of real-valued additive white Gaussian noise (AWGN) samples with $\mathcal{C}\mathcal{N} (0,\frac{N_0}{2})$ values in order to model shot and thermal noise at the receiver.

The ML detector selects the codeword matrix that satisfies
%\cite{mesleh-haas-ahn-yun}

\begin{equation}
  \begin{aligned}
  \label{eq:received-stsk-rf-detection-ml}
     \boldsymbol{\hat{S}} = \displaystyle\arg \min_{\boldsymbol{S}}  \; ||\boldsymbol{Y} - \boldsymbol{H} \boldsymbol{S} ||^{2},
  \end{aligned}
\end{equation}
which implies the codeword matrix which has the closest Euclidean distance with the received codeword matrix.

\subsection{Concatenated Permutations for STSK}
In order to increase the data rate of the PM MIMO scheme, multiple permutation codewords are combined to activate more than one LED at each transmit time. First, we denote $\boldsymbol{c}^{\overline{w}}_{q}$ as the codeword in codebook $\overline{w} \; (1 \leq \overline{w} \leq w)$ at row $q$ and $\boldsymbol{P}_q^{\overline{w}}$ is the equivalent permutation matrix. The concatenated codewords $(\boldsymbol{c}^{1}_{q} \boldsymbol{c}^{2}_{q} \cdots \boldsymbol{c}^{w}_{q})$ are chosen such that the Hamming distance between any two codewords at row $q$ of the codebooks  
\begin{equation} 
    \label{eq:minimum-distance-concatenated-codewords}
    \begin{aligned} 
        d_{\text{m}}(\boldsymbol{c}_q^1, \boldsymbol{c}_q^2) = L.
  \end{aligned}
\end{equation}
This means any $w$ concatenated codewords differ in as many places as the length of a codeword. 
The number of codewords with Hamming distance $L$ from a given codeword $\boldsymbol{c}^{\overline{w}}_{q} \in \boldsymbol{C}_w$ is \cite{frankl-deza}
\begin{equation} 
    \label{eq:minimum-distance-codewords-at-L}
    |K| = L! \cdot \sum_{k=0}^{L} \frac{(-1)^k}{k!},
\end{equation}
and the codewords can be concatenated in order to produce unique codeword transmit matrices. Each matrix has a weight $w$ on the columns and rows, same as the number of concatenated codewords. The transmitted matrix of the concatenated codewords is given as
\begin{equation} 
    \label{eq:matrix-from-concatenated-codewords}
    \boldsymbol{P}_q = \boldsymbol{P}_q^1 + \boldsymbol{P}_q^2 + \dots + \boldsymbol{P}_q^w.
\end{equation} 
In the combined codebook $\boldsymbol{P}_q$, any two codewords ($\boldsymbol{P}_q^{1}, \boldsymbol{P}_q^{2}$) are the same if $\boldsymbol{P}_q^{1} - \boldsymbol{P}_q^{2} = 0$, even if produced by different combined codewords. Hence, only one of the codewords is chosen to avoid duplicate entries.

A simple construction of codewords which satisfy (\ref{eq:minimum-distance-concatenated-codewords}) is achieved by selecting from $L$-order Latin squares \cite{euler}, each consisting of an $L \times L$ integer matrix, derived from a set of $L$ codewords. In each matrix, an integer occurs only once in each row and column and each row and column is a permutation. The permutations that form the Latin square can be constructed for example, by performing a cyclic shift on a codeword. % which moves the positions of the integers to the left and places the first element in the vector to the last element. 
For example, in codebook $\boldsymbol{C}_1$, a cyclic shift on a codeword $\boldsymbol{c}_1^{1} = (c_1^{1}c_2^{1} \dots c_L^{1})$ produces $\boldsymbol{c}_2^{1} = (c_2^{1} c_3^{1} \dots c_{L}^{1} c_1^{1})$. Assuming $\boldsymbol{c}_1^{1}$ is (2314) and is cyclically shifted to (3142), $\boldsymbol{c}_1^{1}$ and $\boldsymbol{c}_2^{1}$ satisfy (\ref{eq:minimum-distance-concatenated-codewords}) and the concatenated codewords produce a matrix

\[\boldsymbol{P}_q = \left[ \begin{array}{cccc}
0 & 1 & 1 & 0 \\
1 & 0 & 0 & 1\\
1 & 1 & 0 & 0\\
0 & 0 & 1 & 1\end{array} \right],\]
with $w = 2$. This enables $w$ unique LED combinations in each time slot in a block using codewords which belong to one of $w$ codebooks $\{\boldsymbol{C}_1, \boldsymbol{C}_2, \dots \boldsymbol{C}_w\}$ for $1 < w < L$. The message symbols are then mapped onto $w$-tuple codewords $(\boldsymbol{c}^{1}_{q} \boldsymbol{c}^{2}_{q} \cdots \boldsymbol{c}^{w}_{q})$ which belong to the same row $q$ in $\boldsymbol{C}_1, \boldsymbol{C}_2, \dots \boldsymbol{C}_w$ respectively. %Hence, any $w$-tuple codewords in a Latin square can be used to activate $w$ LEDs per time slot. 

The concatenated codewords can be added to the conventional $w = 1$ codewords in order to increase the number of codewords. Hence, the total number of codewords achievable is
\begin{equation} 
    \label{eq:codwords-conbined}
    \begin{aligned} 
      Q = Q_1 + Q_2 + \dots + Q_{L-1},
    \end{aligned}
\end{equation} 
where $Q_w$ is the number of codewords available for a given weight $w$. An example is shown in Table \ref{table:LSTSK-permutation-matrix-modulation-order} where the bits per symbol is increased to $2^{(L+1)}$ without increasing the size of $L$.
The increase in the number of active indices also enables higher order constellation points to be modulated on top of the active LED index, hence increasing the data rate when compared to the $w=1$ permutation-aided scheme. Using (\ref{eq:modulation-times.permutation-matrix}), the transmitted matrix is given as
\begin{equation} 
    \label{eq:assignment-problem-explained-sim}
    \begin{aligned} 
      \sum_{i=1}^{M} s_{ij} =
      \sum_{j=1}^{M} s_{ij} = w \cdot a.
    \end{aligned}
\end{equation} 

\begin{table}
\begin{center}
    \caption {Mapping between modulation scheme and codes with $w = 1$ and $w = 2$.}
  \begin{tabular}{c c l l l}
    \toprule
    \multicolumn{1}{c}{} & \multicolumn{2}{c}{$Q = 32, M = 1$} &
      \multicolumn{2}{c}{$Q = 32, M = 2$} \\
      \textbf{Input Bits} & 
    \textbf{Codewords} & $m$ & \textbf{Codewords} & $m$ \\
      \midrule
    00000 & 1234 & 1 & 1234 & 1 \\
    00001 & 1243 & 1 & 1234 & 2 \\
    00010 & 1243 & 1 & 1234 & 1 \\
    00011 & 1243 & 1 & 1234 & 2 \\
    \vdots & \vdots & \vdots & \vdots & \vdots \\
    11100 & 1234, 2143 & 1 & 1234, 2143 & 1 \\
    11101 & 1324, 2413 & 1 & 1324, 2413 & 2 \\
    11110 & 3241, 4132 & 1 & 3241, 4132 & 1 \\
    11111 & 3124, 4312 & 1 & 3124, 4312 & 2 \\
    \bottomrule
  \end{tabular}
\label{table:LSTSK-permutation-matrix-modulation-order}
\end{center}
\end{table}
Given equiprobable symbols with unique one-to-one mapping onto codewords, the pairwise error probability of receiving a codeword signal $\boldsymbol{\hat{S}}$ when $\boldsymbol{S}$ is transmitted is given as 
\begin{equation}
  \begin{aligned}
  \label{eq:pep-stsk}
    Pr(\boldsymbol{\hat{S}} \rightarrow \boldsymbol{S} | \boldsymbol{H}) \leq \text{Q}\left(\sqrt{\frac{E_s}{2N_0} ||\boldsymbol{H}(\boldsymbol{S} - \boldsymbol{\hat{S}})||_F^2}\right).
  \end{aligned}
\end{equation}
Within a transmit block, the symbol to noise ratio is $\frac{E_s}{N_0} = \frac{(rI)^2}{N_0} T_s$ for sampling time $T_s$ and an optical-to-electrical conversion coefficient $r$.  By using the union bound method, the upper bound of the Bit Error Rate (BER) in (\ref{eq:ber-stsk}) compares all possible $M \cdot Q$ codeword matrices and intensity combinations. The term $\text{d}_{\text{m}}(b_{m_1}^{(q_1)}, b_{m_2}^{(q_2)})$ denotes the Hamming distance between the received bits $b_{m_2}^{(q_2)}$ when actually, $b_{m_1}^{(q_1)}$ was transmitted.
\begin{figure*}[!t]
% ensure that we have normalsize text
\normalsize
% Store the current equation number.
%\setcounter{mytempeqncnt}{\value{equation}}
% Set the equation number to one less than the one% desired for the first equation here.
% The value here will have to changed if equations% are added or removed prior to the place these
% equations are referenced in the main text.
%\setcounter{equation}{6}
    \begin{equation}
    \label{eq:ber-stsk}
        \text{BER} \leq \frac{1}{M \cdot Q \; \text{log}_2(M \cdot Q)} \sum_{m_1=1}^{M}\sum_{q_1=1}^{Q}\sum_{m_2=1}^{M}\sum_{q_2=1}^{Q} \text{d}_{\text{m}}(b_{m_1}^{(q_1)}, b_{m_2}^{(q_2)}) \;\text{Q}\left(\sqrt{\frac{E_b}{2N_0} ||\boldsymbol{H}(I_{m_1}\boldsymbol{S}^{(q_1)} - I_{m_2}\boldsymbol{S}^{(q_2)}})||_F^2\right).
    \end{equation}
    %\begin{equation}
        %\label{eqn_dbl_y}y = 4 + 6 + 8 + 10 + 12 + 14 + 16 + 18 + 20+ 22 + 24+ 26 + 28 + 30
    %\end{equation}
% Restore the current equation number.
%\setcounter{equation}{
 %   \value{mytempeqncnt}
  %  }
% IEEE uses as a separator
\hrulefill
% The spacer can be tweaked to stop underfull vboxes.
\vspace*{4pt}
\end{figure*}
%\subsection{Differential Permutation-Aided STSK Scheme}
%\begin{equation}
%  \begin{aligned}
%  \label{eq:differential-send-pstsk}
%    \boldsymbol{X} = s \boldsymbol{P}_q,
%  \end{aligned}
%\end{equation}

%The $\boldsymbol{S}$ is differentially encoded and transmitted as
%\begin{equation}
%  \begin{aligned}
%  \label{eq:differential-send-pstsk}
%    \boldsymbol{S} = \boldsymbol{S}(t-1) \boldsymbol{X},
%  \end{aligned}
%\end{equation}
%The ML detection is
%\begin{equation}
%  \begin{aligned}
%  \label{eq:differential-send-pstsk}
%    \boldsymbol{\hat{X}} = \displaystyle\arg \min_{q,l}}  \; ||\boldsymbol{\hat{Y}} - %\boldsymbol{\hat{Y}}(i-1)\Tilde{X}} ||^{2}
%  \end{aligned}
%\end{equation}

\section{Proposed Optimization Algorithms for Soft-detection Detection}
From the received matrix $\boldsymbol{Y} \in [y_{ij}]$ at each block, the detector processes $\hat{\boldsymbol{Y}} \in [\hat{y}_{ij}]$ such that $\hat{y}_{ij} = -y_{ij}$. For $w = 1$, the decoder finds the corresponding codeword $\hat{\boldsymbol{c}}$ at row $q$ that produces the cost
\begin{equation} 
   \label{eq:assignment-problem-solution-1}
    g_{1} = \sum_{i=1}^{L}\sum_{j=1}^{L} \hat{y}_{ij} s_{ij},
\end{equation}
that minimizes (\ref{eq:assignment-problem-solution-1}).
For $1 < w < L$, the decoder finds the set of $w$-tuple codewords $\{\hat{\boldsymbol{c}}^{1}_{q}, \hat{\boldsymbol{c}}^{2}_{q}, \cdots, \hat{\boldsymbol{c}}^{w}_{q}\}$ which belong to the same row $q$ in \{$\boldsymbol{C}_1, \boldsymbol{C}_2, \dots, \boldsymbol{C}_w \}$ that produce the cost

\begin{equation}
    \begin{split}
g_{1} = \arg \min_{\boldsymbol{c}_{q} \in C} (\sum_{i=1}^{L}\sum_{j=1}^{L} \hat{y}_{ij} s^{(\boldsymbol{c}^{1}_{q})}_{ij} + \sum_{i=1}^{L}\sum_{j=1}^{L} \hat{y}_{ij} s^{(\boldsymbol{c}^{2}_{q})}_{ij} + \dots \\ + \sum_{i=1}^{L}\sum_{j=1}^{L} \hat{y}_{ij} s^{(\boldsymbol{c}^{w}_{q})}_{ij}), \quad \mbox{for} \ q = 1, 2,\dots, Q,
    \end{split}
   \label{eq:assignment-problem-multiple-solution}
\end{equation}
that minimize (\ref{eq:assignment-problem-multiple-solution}). Here, $s^{(\boldsymbol{c}^{\overline{w}}_{q})}_{ij}$ values are the elements in $\boldsymbol{S}$, produced by a codeword at row $q$. Each codeword $\boldsymbol{c}^{1}_{q}, \boldsymbol{c}^{2}_{q}, \dots, \boldsymbol{c}^{w}_{q}$ belongs to codebooks $\boldsymbol{C}_1, \boldsymbol{C}_2, \dots, \boldsymbol{C}_w$ respectively. %In the case where $M > 1$, the output of the soft-decision decoder is compared with all possible 

\subsection{Brute Force Soft-Decision Receiver}
The brute force (BF) of the soft-decision decoders solves
\begin{equation} 
    %Z_1 = \sum _{i=1}^{M} \sum _{j=1}^{M} y_{ij} x_{ij}, \\ 
    g_{\text{BF}} = \arg \min_{\boldsymbol{c}_{q} \in C} (\sum_{i=1}^{L}\sum_{j=1}^{L} \hat{y}_{ij} s^{(\boldsymbol{c}_{q})}_{ij}), \quad \mbox{for} \ q = 1, 2,\dots Q,
   \label{eq:assignment-problem-solution-permutation-ml}
\end{equation}
where $s^{(\boldsymbol{c}_{q})}_{ij}$ is each element in the permutation matrix produced by a codeword $\boldsymbol{c}$. The BF SD receiver ranks the costs $g_1, g_2, \dots, g_{Q}$ and chooses the lowest cost $g_{\text{BF}}$ that produces the permutation matrix $\boldsymbol{P}_{\text{BF}}$ and whose corresponding codeword $\hat{\boldsymbol{c}}_{\text{BF}} \in C$.

\subsection{Branch and Bound}
\begin{figure}
    \centering\includegraphics[width=20pc]{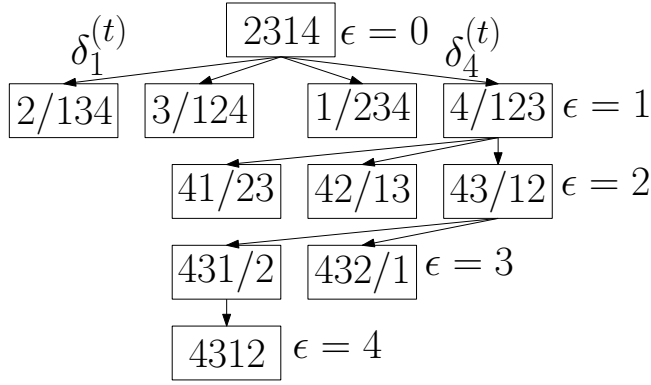}
    \caption{Tree-based method to decode permutation codes using branch and bound with $L = 4$.}
    \label{fig:branch-and-bound-tree-diagram}
\end{figure}
The branch and bound algorithm (BB) \cite{little-murty-sweeney-karel} can be used to solve a $w = 1$ received matrix in ($\ref{eq:assignment-problem-solution-1}$) using the tree-based method. $L+1$ levels defined as $\epsilon = 0, 1, \dots, L$ are created. Each node has an associated cost $\delta_{\epsilon}^{(t)} = \delta_{\epsilon}^{(1)}, \delta_{\epsilon}^{(2)}, \dots, \delta_{\epsilon}^{(L-\epsilon+1)}$. The initial node at $\epsilon = 0$ is any of the codewords used at the encoder. The surviving node at $\epsilon = 1$ is determined by finding the node that satisfies
\begin{equation} 
%    \hat{\delta}_{\epsilon}^{(t)} = \min_{1 \leq t \leq M} \sum_{i=1}^{M} \hat{y}_{it}, \:\:\: \;\;\;\; \epsilon = 1. %t = 1, 2,\dots,M. %1 \leq t \leq M. t = 1, 2,\dots,M.
    \hat{\delta}_{\epsilon}^{(t)} = \min_{1 \leq t \leq L} (\hat{y}_{\epsilon t} + \sum_{i,j} \hat{y}_{ij}),  %t = 1, 
    \label{eq:branch-and-bound-level-0}
\end{equation}
for $\epsilon = 1, 2 \leq i \leq L, 0 < j \leq L: j \neq t$.
The surviving node $\hat{t}_{\epsilon}$ which satisfies (\ref{eq:branch-and-bound-level-0}) is used to create $L-\epsilon$ branches for levels $\epsilon = 2$. The surviving node at each levels $\epsilon > 1$ is evaluated as
\begin{equation} 
%    \hat{\delta}^{(t)}_{\epsilon} = \min_{\substack{1 \leq t \leq M-\epsilon+1, \\ \Tilde{y} = \infty} } (\Tilde{y} + \sum_{i=\epsilon}^{M} \hat{y}_{it}), \:\;\;\; 2 \leq \epsilon \leq M,
    %\hat{y}_{i\hat{\delta}_{\epsilon_2}} = \infty. %t = 2,\dots,M, \; \hat{y}_{i\hat{\delta}_{\epsilon_2}} = \infty.
    \hat{\delta}_{\epsilon}^{(t)} = \min_{1 \leq t \leq L} (\hat{y}_{(\epsilon-1)\hat{t}_{(\epsilon-1)}} + \hat{y}_{\epsilon t} + \sum_{i,j} \hat{y}_{ij}),  %t = 1 \:\: i = \epsilon+1, \dots, M, j = 1, \dots, M, 
    \label{eq:branch-and-bound-level-1}
\end{equation}
for $\epsilon > 1, \epsilon+1 \leq i \leq L, 1 \leq j \leq L : j \neq t, j \notin \boldsymbol{\hat{t}}$. Note that $\boldsymbol{\hat{t}}$ is a vector containing the surviving nodes and (\ref{eq:branch-and-bound-level-1}) is repeated until the $L - 1$-th node to solve the permutation codeword.

\subsection{Iterative Soft-Decision Detection}
The iterative decoder finds the maximum cost of $\hat{\boldsymbol{Y}}$ that produces a codeword $\hat{\boldsymbol{c}}^{n}_{q}$ in each codebook $\{\boldsymbol{C}_1, \boldsymbol{C}_2, \dots, \boldsymbol{C}_w\}$ at each iteration $e$. Assuming $\hat{\boldsymbol{c}}^{1}_{1}$ is the decoded codeword for $\boldsymbol{C}_1$ and $\hat{\boldsymbol{c}}^{2}_{2}$ is decoded for $\boldsymbol{C}_2$, then the decoder chooses between the codeword pairs \{$\hat{\boldsymbol{c}}^{1}_{1}, \hat{\boldsymbol{c}}^{2}_{1}$\} and \{$\hat{\boldsymbol{c}}^{1}_{2}, \hat{\boldsymbol{c}}^{2}_{2}$\} having the highest cost.

At iteration $e = 1$, the Hungarian algorithm finds the minimum cost $g_1$ using steps described in \cite{kuhn}. This produces a row-column pair ${\{(1, \hat{c}_1), (2, \hat{c}_2), \dots, (Q, \hat{c}_{L})}\}$ which corresponds to a codeword $\boldsymbol{\hat{c}} = (\hat{c}_1 \hat{c}_2 \dots \hat{c}_L$) with the minimum cost $g_1$. If $\boldsymbol{\hat{c}} \notin \{\boldsymbol{C}_1, \boldsymbol{C}_2, \dots, \boldsymbol{C}_w\}$, then solutions at $e = 2, 3, \dots, Q$ are found using Murty's \cite{murty}. The solution matrix from the Hungarian algorithm at $e = 1$ is used to create $L-1$ nodes or subsets $U_1, U_2, \dots, U_{L-1}$. The nodes are created by partitioning the solution matrix at $e = 1$ such that $U_1 = {\{\overline{(1, j_1)}}\}, \dots , U_{L-1} = {\{(1, j_1), \dots, \overline{(L-1, j_{L-1})}}\}$.
The items with a bar ($\overline{\cdot,\cdot}$) in $U_1, U_2, \dots, U_{L-1}$ are replaced with $\infty$ while the other items are excluded. By solving for the minimum cost of each derived node, the next assignment is then derived from the node with the lowest cost. This process can be iterated from $e = 2, 3, \dots, Q$ until $\hat{\boldsymbol{c}} \in \{\boldsymbol{C}_1, \boldsymbol{C}_2, \dots, \boldsymbol{C}_w\}$.

\begin{figure}
\centering\includegraphics[width=22pc]{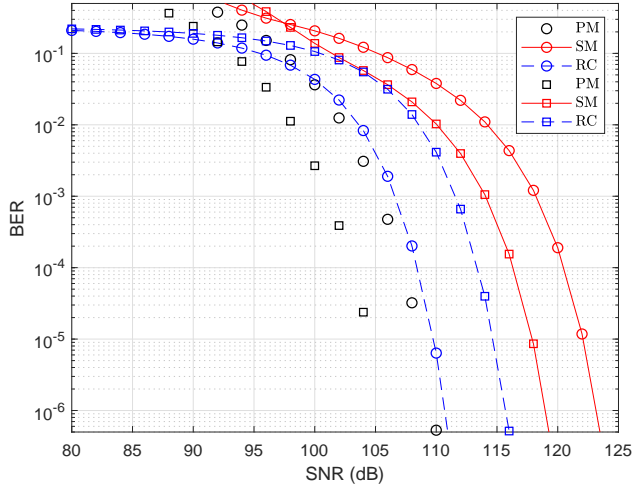}
\caption{BER comparisons of PM with RC and SM for two channel matrices transmitting 4 bits. The circles and squares are plots for $\boldsymbol{H}_{0.2}$ and $\boldsymbol{H}_{0.6}$ respectively.}
\label{fig:ber-comapring-rc-sm-pa}
\end{figure}

\begin{figure}
\centering\includegraphics[width=22pc]{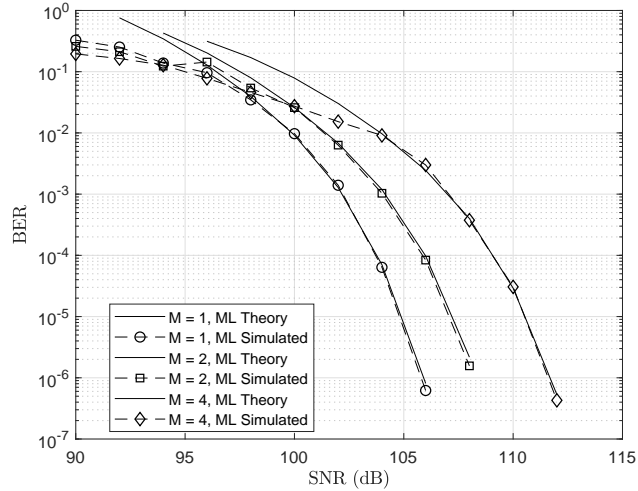}
\caption{BER of concatenated codebooks of $w = 1$ and $w = 2$ containing 24 and 8 codewords respectively.}
\label{fig:ber-concatenated-4x4-32cw-w1-2}
\end{figure}

\section{Simulation Results}
The error rate performance of the space-time schemes are evaluated for an indoor environment. The transmitter units are assumed to be placed at the top of the room while the receiver units are placed on a table in the room, 1.75 m from the transmitters. The number of LEDs in the transmitter and PDs in the receiver are both four and the channel gain between each transmitter and receiver are calculated using (\ref{eq:channel-matrix}). Spacing between the receivers is 0.1 m while different transmitter spacing $D_\text{tx}$ are considered to generate the channel matrix $\boldsymbol{H}_{D_{\text{tx}}}$ used for the numerical simulations. The channel matrix for $D_\text{tx} = 0.2$ m also adopted from \cite{fath-haas}

\begin{equation*} 
  \begin{aligned}
      \boldsymbol{H}_{0.2} = \left[ 
      \begin{array}{cccc}
            1.0708 & 0.9937 & 0.9937 & 0.9226 \\
            0.9937 & 1.0708 & 0.9226 & 0.9937\\
            0.9937 & 0.9226 & 1.0708 & 0.9937\\
            0.9226 & 0.9937 & 0.9937 & 1.0708 
        \end{array} \right] \times 10^{-4},
  \end{aligned}
  \label{eq:channel-matrix}
\end{equation*}
with values obtained using (\ref{eq:vlc-tx-h}) with 0.2m and 0.1m spacing between the transmitters and receivers respectively. $A_{\text{pd}}$ is assumed to be unity while both $\Phi_{\frac{1}{2}}$ and $\Psi$ are set to $15^{\circ}$ \cite{bouchet-faulkner-grobe-gueutier-langer-nerreter, brien-falkner-bouchet-taback-wolf}. We shorten the simulation properties by describing the schemes using P($L, Q, M, w$) for single weight codebooks and P($L, Q, M, \{w_1, w_2\}$) for multiweight codebooks.

Fig. (\ref{fig:ber-comapring-rc-sm-pa}) shows the BER using two transmitter spacings $D_\text{tx} = 0.2$ m and $D_\text{tx} = 0.6$ m. The advantage of the space-time code over spatial multiplexing schemes such as RC and SM is seen in the BER performance as well as better transmitter spacing improving, the PM's performance. In Fig. \ref{fig:ber-concatenated-4x4-32cw-w1-2}, the combination of permutation codewords to increase the data rate is shown with the simulations matching the derived bound in (\ref{eq:ber-stsk}). In the SD decoder of the combined codebooks, it is assumed the receiver can differentiate between transmitted $w = 1$ and $w > 2$ matrices.

%\begin{figure}
%\centering\includegraphics[width=22pc]{diagrams/results/pa_vlc_m1_q16_h02_snr90_2_106.eps}
%\caption{Comparison of soft-decision techniques with ML performance for P(4, 16, 1, 1).}
%\label{fig:ber-4x4-16cw-w1}
%\end{figure}

\begin{figure}
\centering\includegraphics[width=22pc]{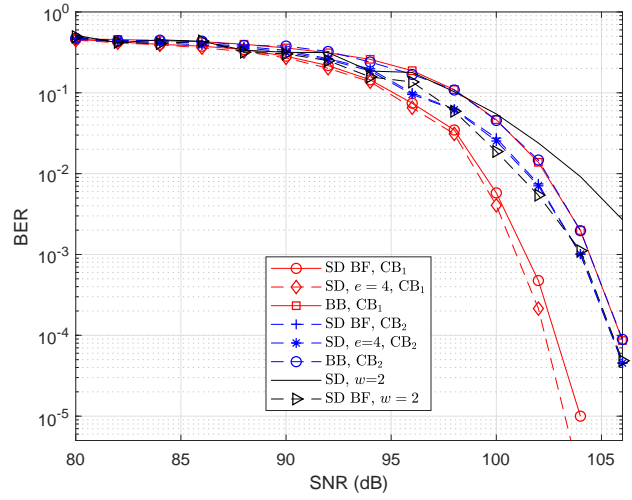}
\caption{Comparison of soft-decision techniques with ML performance for P(4, 8, 1, 1) and P(4, 8, 1, 2) with $\boldsymbol{H}_{0.2}$.}
\label{fig:ber-4x4-8cw-w1}
\end{figure}

%\begin{figure}
%\centering\includegraphics[width=22pc]{diagrams/results/mols_vlc_m4_q8_h02_w_2_snr90_2_110.eps}
%\caption{Comparison of soft-decision decoders with ML decoder for P(4, 8, 4, 2) scheme with $w = 2$.}
%\label{fig:ber-4x4-8cw-w2}
%\end{figure}

\subsection{BER Analysis}
The BER of the ML performance of the scheme with perfect knowledge of the CSI is compared with the BER of the SD decoders. For the scheme using the concatenated codebook, equal total transmit power is maintained for all codewords by transmitting intensity $\frac{I_m}{w}$ over each active transmitter.

\begin{figure}
\centering\includegraphics[width=22pc]{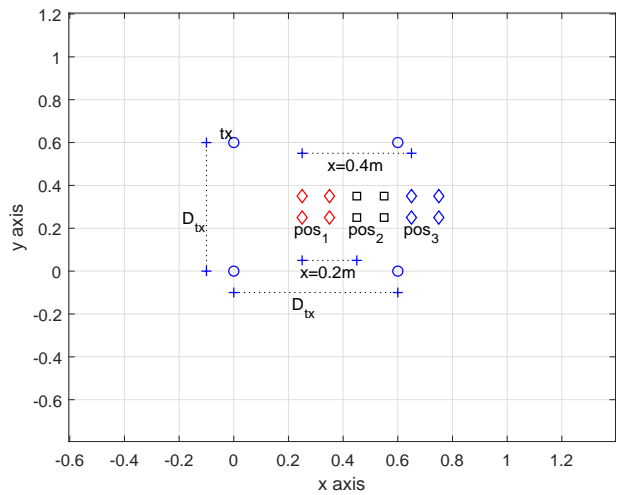}
\caption{Top view cross-section of transmitter and receiver placements.}
\label{fig:tx_rx_placements}
\end{figure}

In the results, the SD decoders decode the channel output directly, hence the CSI is not required.
In Fig. \ref{fig:ber-4x4-8cw-w1} where $w = 1$, appropriate codeword selection can improve the performance of the SD decoder rather than the minimum distance of the code. Hence, choosing codewords which exploit the channel gain properties of the $\boldsymbol{H}$ matrix can produce a different performance of the SD decoder. The effect of codeword selection is shown using two different codebooks transmitting the same rate of information. The two codebooks are $\text{CB}_1$ = (4321, 4132, 3124, 3412, 2431, 2143, 2314, 1342) and $\text{CB}_2$ = (1234, 2134, 2143, 3214, 3124, 3241, 1342, 1432). $\text{CB}_1$ is chosen such that the lower channel gains such that for each codeword, the total channel gain
\begin{equation} 
   \label{eq:codeword-cost}
    \sum_{i=1}^{L}\sum_{j=1}^{L} h_{ij} s_{ij},
\end{equation}
is minimal. $\text{CB}_2$ is the opposite where the codewords target the higher channel gains. It can be seen in Fig. \ref{fig:ber-4x4-8cw-w1} that the codebook which exploits the lower channel gains produces better BER performance from SD decoding. The BB decoder performance is however the same for both codebooks. For $w=2$, the BER plots are compared with $w=1$ with same power allocation to activated LEDs and same bits per transmit block. However, codeword selection is limited due to the number of codewords that meet the criteria in (\ref{eq:minimum-distance-concatenated-codewords}). The SD decoder of $w=2$ codewords differ in about 3dB from the ML while the SD which uses the BF detection matches the ML BER performance.

\subsection{Mobile Receiver without CSI}
The performance of the SD decoder as the receiver moves away from the direct LOS of the transmitter units is shown in Fig. \ref{fig:ber-concatenated-4x4-32cw-w1-2} for combined codebooks of $w=1$ and $w=2$. The transmitter and receiver setups are shown in Fig. \ref{fig:tx_rx_placements} with the blue circles showing the position of the transmitter (tx) with $D_\text{tx} = 0.6$ m. The first position ($\text{pos}_1$) of the receiver is centered with the transmitter while the second and third positions ($\text{pos}_2$ and $\text{pos}_3$) are moved $x$ m away from the first position. Perfect synchronization between the transmitters and receivers is assumed and three different positions at $x = 0.0, 0.2$ and $0.4$ m are shown in clusters of four diamond and black square shapes. Combining these two codebooks increases the bits per block for the same time block and transmitter units. Since the decoder does not require CSI, the performance degrades from the ML with perfect CSI but is able to decode with about 4dB difference when the receiver moves 0.2 m away from the transmitters' LOS. At $x = 0.4$ m away from the transmitter, the SD fails to decode correctly.

\begin{figure}
\centering\includegraphics[width=22pc]{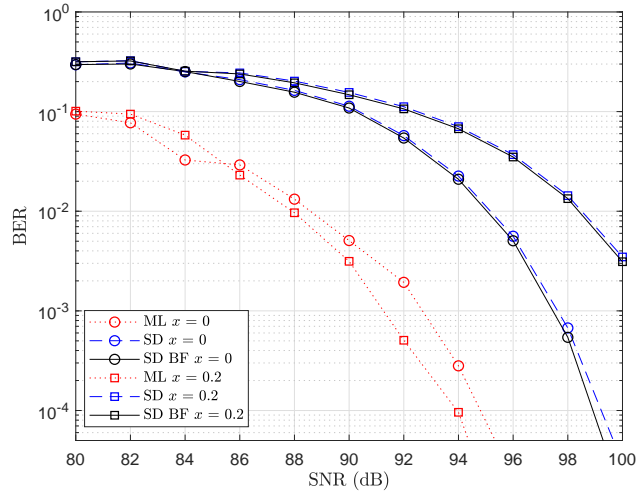}
\caption{BER of concatenated codebooks of P(4, 8, 1, \{1, 2\}) using $\Tilde{\boldsymbol{H}}_{0.6}$ for different positions of the receiver.}
\label{fig:ber-w1-2-link-blockage-h06}
\end{figure}

Link blockage sets up the transmitter and receiver such that a PD is removed from the FOV of an LED. This has shown to assist with improving BER performance by cancelling possible interference caused by the link. Using the $\boldsymbol{H}_{0.6}$ channel matrix as an example, the link blockage between transmitter and receiver pairs (1, 4), (2, 3), (3, 2) and (4, 1) produces a matrix
\begin{equation*} 
  \begin{aligned}
      \Tilde{\boldsymbol{H}}_{0.6} = \left[ 
      \begin{array}{cccc}
            0.6888 & 0.5559 & 0.5559 & 0.0000 \\
            0.5559 & 0.6888 & 0.0000 & 0.5559\\
            0.5559 & 0.0000 & 1.0708 & 0.5559\\
            0.0000 & 0.5559 & 0.5559 & 0.6888 
        \end{array} \right] \times 10^{-4},
  \end{aligned}
  \label{eq:channel-matrix-h06-link-blockage}
\end{equation*}
with the zero elements showing the removal of possible channel gains between the corresponding transmitter and receiver. The BER improvement is shown in Fig. \ref{fig:ber-w1-2-link-blockage-h06} with the SD decoder matching its OD performance but with 4dB loss when compared with the ML decoder.
\begin{figure}
\centering\includegraphics[width=22pc]{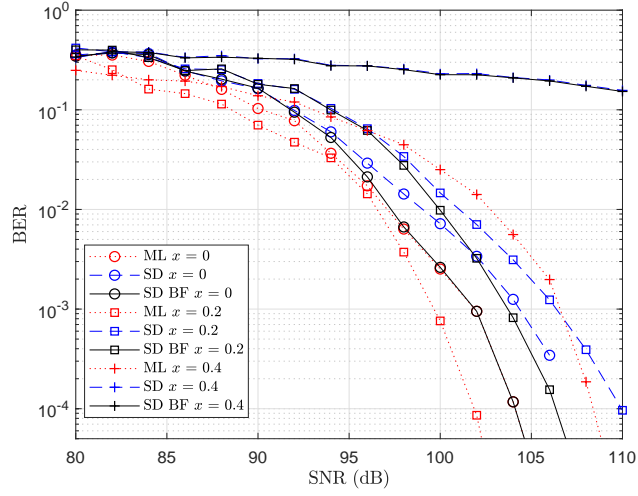}
\caption{BER of concatenated codebooks of P(4, 8, 1, \{1, 2\}) transmitting 5 bits with receiver at the transmitter's center using  $\boldsymbol{H}_{0.6}$ and different positions of the receiver.}
\label{fig:ber-w1-2-moving-rx-h06}
\end{figure}

\subsection{Data Rate Analysis}
%\subsection{Data Rate Analysis}
%Since $\text{log}_2(M \cdot Q)$ bits per symbol can be transmitted per block, only a maximum of $2^{(\text{log}_2Q)}$ codewords can be selected from the total available $L!$ codewords when $w = 1$. 
%\begin{table}
%\begin{center}
%    \caption {Achievable number of codewords for different values of $w$ given a codeword length.}
%  \begin{tabular}{ccSSSSS}
%    \toprule
%    \multicolumn{1}{c}{} & \multicolumn{6}{c}{$w$} &
%      \multicolumn{5}{c}{} \\
%      \multirow{$L$} & 
%    \multirow{1} & {2} & {3} & {4} & {5} & {6} \\
%      \midrule
%    3 & 6 & 2 & & & & \\
%    4 & 24 & 90 & 24 & & & \\
%    5 & 120 & 2040 & & & & \\
%    6 & 720 & 336 & & & & \\
%    7 & 5040 & 2215 & & & & \\
%    8 & 40320 & 226 & & & & \\
%    \bottomrule
%  \end{tabular}
%\label{table:acheivable-rates}
%\end{center}
%\end{table}

The combination of codebooks with different weights makes more codewords available in order to increase the bits per symbol. Consider the codebook with $L = 4$ which has 24 ($L!$) matrices for $w = 1$. An exhaustive search of unique codewords that satisfy (\ref{eq:minimum-distance-concatenated-codewords}) produces 90 matrices for $w = 2$. For the same $L$, 24 matrices are found for $w = 3$. If the codebooks with the 3 weights are combined, 7 bits per symbol can be transmitted by each matrix compared to the possible 4 bits per symbol achievable for the $w = 1$ codebook.%Table \ref{table:acheivable-rates} lists the possible number of unique concatenated codwords that are achievable for a given $w$.

\subsection{Complexity Analysis of Soft-Decision Decoders}
%The BB decoder in Fig. \ref{fig:ber-4x4-16cw-w1} uses a complexity of $O(L^3 \text{ln}^2(L))$ while the iterative soft-decision decoder in Figs. \ref{fig:ber-4x4-16cw-w1} and \ref{fig:ber-4x4-8cw-w2} uses a worst-case complexity of $O(L^3)$ \cite{cox-miller-danchick-newnam}. The BF soft-decision decoder in Figs. \ref{fig:ber-4x4-16cw-w1}-\ref{fig:ber-concatenated-4x4-32cw-w1-2} operates a $O(L \cdot Q!)$ \cite{smith} complexity in order to find the optimal cost that corresponds to the decoded codeword.

The decoders' complexities are analysed based on the different operations required in the decoding process while excluding circuitry complexities. Considering the input size  $L \times L$ into the decoder which represents a received coded symbol. The ML decoder compares the received codeword matrix with all possible $Q$ codeword matrices used at the transmitter in order to make a decision. Depending on the method used, each element in the received codeword matrix can be compared with each element in the likely transmitted codeword matrix in order to find the distance between the two matrices. This method will additionally require $L^2$ operations. Hence the ML decoder's complexity can be approximated as $O(Q \cdot L^2)$. With the same $L \times L$ input size, the algorithms in the SD decoders perform their respective decoding operations. Using the tree-based method for BB, each level requires $L$ operations i.e. for each node, and each node computes the individual costs before finding the minimum cost at each level. The complexity of the BB algorithm can then be approximated as $O(L^3\text{log}L)$ since the number of nodes to compute reduce at each level. In the SD decoder, each term in \ref{eq:assignment-problem-multiple-solution} requires a complexity of $O(L^3)$ at $e=1$ and \cite{smith} $O(L^4)$ \cite{liu-shell} at $e>1$. Therefore, the worst-case complexity can be approximated to $O(L^4)$. This implies that the complexity of the SD decoder is dependent on the size of $L$ rather than the size of the codebook $Q$. In large codebooks which can assist with creating codeword matrices for higher rate data, $Q$ \(\gg\) $L$. The SD BF method finds the cost of each codeword and finds the codeword that minimizes (\ref{eq:assignment-problem-solution-permutation-ml}). This approximates the complexity to $O(L \cdot Q)$ in order to find the optimal cost that corresponds to the decoded codeword.

%For each received coded signal at the output of the channel, the TD requires a complexity of $O(M^2)$ to decide if each sample in the $M \times M$ matrix is above or below the threshold. For the PSDD, the complexity of the iterative decoder in Scheme 1 and 2 is $O(M^3)$ at $g = 1$ and $O(M^4)$ for $2 \leq g\leq M$. When $g > 1$, the complexities sum up, hence a worst-case complexity of $O(M^4)$ \cite{miller-stone-cox}. At each level of the tree-based BB method in Fig \ref{fig:branch-and-bound-tree-diagram} and (\ref{branch-and-bound-level-0}), $M$ different computations are required on all the $M$ nodes in order to find the minimum cost at each level. Since the number of nodes to process reduce at each level, the worst-case complexity is considered and can be approximated as $O(M^3\text{log}M)$ \cite{lageweg-lenstra-rinnooy-kan}. Schemes 2 and 4 require a lookup process to demap to the binary equivalent at a complexity of $O(M \text{log}M)$. While the PSDD schemes eliminate the complexity of ED and TD, they increase the complexity in comparison with ED and TD.

\section{Conclusion}
Concatenated permutations are used to increase the data rate of the PM-aided MIMO scheme in VLC by combining permutation matrices of different weights. Low-complexity, SD techniques are also used to detect the transmitted signals without the knowledge of the CSI. The results show the soft-decision decoder can match the ML decoder in some codebooks and the decoding complexities are also analyzed. In future work, low-complexity code construction techniques are required because the brute force code construction is prohibitive computationally. Also, low-complexity, SD techniques which can decode without CSI knowledge, yet with closer BER performance will assist with improving the overall system.

\section*{Acknowledgement}
The authors would like to thank and acknowledge the financial support provided by South Africa's National Research Foundation (112248 \& 114626) and the Sibanye-Stillwater Digital Mining Laboratory (DigiMine), Wits Mining Institute (WMI), University of the Witwatersrand, Johannesburg, South Africa.

\bibliography{main}
\bibliographystyle{elsarticle-num}

\end{document}